\documentclass[prd,a4paper,aps,showpacs,twocolumn,floats]{revtex4}
\usepackage{epsfig}
\usepackage{amssymb}

\newcommand{\bx}{{\mathbf x}}

\newcommand{\al}{\alpha}

\newcommand{\de}{\delta}

\newcommand{\La}{\Lambda}

\newcommand{\Om}{\Omega}

\newcommand{\lap}{\triangle}

\newcommand{\be}{\begin{equation}}
\newcommand{\ee}{\end{equation}}
\newcommand{\bea}{\begin{eqnarray}}
\newcommand{\eea}{\end{eqnarray}}
\newcommand{\bean}{\begin{eqnarray*}}
\newcommand{\eean}{\end{eqnarray*}}
\newcommand{\dd}{\partial}

\begin{document}

\title{
No-go theorem for k-essence dark energy}
\author{Camille Bonvin}
\email{camille.bonvin@physics.unige.ch}
\affiliation{D\'epartement
de Physique Th\'eorique, Universit\'e de
  Gen\`eve, 24 quai Ernest Ansermet, CH--1211 Gen\`eve 4, Switzerland}
\author{Chiara Caprini}
\email{chiara.caprini@physics.unige.ch}
\affiliation{D\'epartement de Physique Th\'eorique, Universit\'e
de Gen\`eve, 24 quai Ernest Ansermet, CH--1211 Gen\`eve 4,
Switzerland}

\author{Ruth Durrer}
\email{ruth.durrer@physics.unige.ch}
\affiliation{D\'epartement de
Physique Th\'eorique, Universit\'e de
Gen\`eve, 24 quai Ernest Ansermet, CH--1211 Gen\`eve 4, Switzerland}

\date{\today}

\begin{abstract}
We demonstrate that if k-essence can solve the coincidence problem and
play the role of dark energy in the
universe, the fluctuations of the field have to propagate
superluminally at some stage. We argue 
that this implies that successful k-essence models violate
causality. It is not possible to define a time ordered succession of
events in a Lorentz invariant way. Therefore, k-essence cannot arise as low
energy effective field theory of a causal, consistent high energy theory. 

\end{abstract}

\pacs{98.80,11.30.Cp}

\maketitle

\label{sec:intro}

Cosmological observations indicate that the expansion of the Universe is
presently in an accelerating phase~\cite{sn}. In a homogeneous and isotropic
universe, this can be obtained, if the energy density is dominated by
a component $x$ with $w_x =P_x/\rho_x < -1/3$; here $\rho_x$ is the
energy density of the component $x$ and $P_x$ is its pressure. The
simplest example of such a component which is compatible with
observations is a cosmological constant of the order of $\La
\simeq 2H_0^2$, where $H_0$ is the present value of the Hubble
parameter. Apart from the smallness of this value which cannot be
explained by any sensible theory of fundamental interactions, it is
perturbing that the value of the cosmological constant should just be
such that it comes to dominate today.

In order to alleviate this coincidence problem quintessence~\cite{qu1}
and k-essence~\cite{ke1} have  been proposed. In these models a scalar
field has the property that at early times, in the radiation dominated
universe, its energy density 'tracks' the one of the cosmic fluid and
therefore naturally provides a sizeable fraction of the energy
density of the universe. Quintessence also tracks the matter energy 
density during matter domination, but the mechanism which leads to the
domination of quintessence today is not clearly identified.

In the case of k-essence the situation is different. 
Within a certain range of initial conditions, the energy density sharply drops
after the beginning of the  matter dominated era and assumes an
equation of state $P_k \simeq -\rho_k$ (de Sitter phase). Afterwards its
contribution can either rise to
dominate the energy density with an equation of state of $w_k =$
constant $ < 0$, or become comparable to that of
matter and start to 'track' the matter. The radiation
tracker, de Sitter phase and k-essence domination (or matter tracker) are all 
attractor solutions of the k-essence evolution equation. The k-essence field
is driven from one to the other by the evolution of the
universe~\cite{ke1,ke+,mal}.   
This looks promising as a solution to the coincidence
problem~\footnote{In Ref.~\cite{mal} it is, however, argued that the
  basin of attraction of the examples of Refs.~\cite{ke1,ke+}
  is disturbingly small.}.

However, in this letter we shall show that a k-essence field which behaves 
in the way described above cannot emerge as the low energy limit of a 
consistent, causal high energy  theory (be this a quantum field
theory or string theory, see Ref.~\cite{AA}). 

The k-essence model is characterized by non-standard kinetic energy
terms \cite{ke1,ke+}. The problem of
acausalities in scalar field theories with non-quadratic Lagrangian
has also been addressed in Ref.~\cite{bek}.   
The action of k-essence is given by 
\be\label{action}
 S = \int d^4x \sqrt{-g}\left[- \frac{R}{6} + P(\phi,X)\right]~,
\ee
were $\phi$ is the k-essence field and
$X=\frac{1}{2}\nabla_\mu\phi\nabla^\mu\phi$. 
We use units with $\frac{8\pi G}{3}=1$ and the metric signature is
$(+,-,-,-)$.  Furthermore, one assumes that the Lagrangian can be factorized  
$P(\phi,X) =K(\phi)p(X)$ with $K(\phi)>0$. A standard scalar field with $ P(\phi,X)=X$ does
of course not have the behaviour we are looking for: one therefore allows
$p$ to be an arbitrary, monotonically growing function of $X$ \cite{ke+}.

Varying the above action with respect to the metric one obtains the
energy momentum tensor,
\bea\label{em}
T_{\mu\nu} &=& 
  \frac{\dd P(\phi,X)}{\dd X}\nabla_\mu\phi\nabla_\nu\phi -
  P(\phi,X)g_{\mu\nu}\\
 &=& (\rho_k +P_k)u_\mu u_\nu  -P_kg_{\mu\nu}
\eea
with $u_\mu =(2X)^{-1/2}\nabla_\mu\phi$, $\rho_k = 2X \frac{\dd
  P}{\dd X} -P$ and $P_k=P$ 
Note that in a homogeneous and isotropic universe
  $\nabla_\mu\phi$ is time-like, $X>0$.  

The idea is of course, that $\phi$ is a low energy effective degree of freedom
of some fundamental high energy theory \cite{ke1,ke+,ke2} which
should satisfy basic criteria: among them, most importantly, 
Lorentz invariance and causality. No
information should propagate faster than the speed of light $c=1$.
Let us translate this basic requirement to the low energy effective
degree of freedom $\phi$. We consider the cosmic background
solution with small fluctuations, $\phi = \phi_0(t)
+\de\phi(t,\bx)$. It is then easy to derive an equation of motion for
the fluctuations $\de\phi$ which is of the generic form~\footnote{In
  principle we would have to write the evolution equation for a gauge
  invariant quantity, like the Bardeen potential, but it has the same
  form and, more importantly, the same velocity of propagation, $c_s^2$.}
\bea\label{fluc}
&&\ddot{\de\phi} + \al\dot{\de\phi} + \beta\de\phi +c_s^2\lap\de\phi =0
\quad \mbox{ where } \quad \\
&&\lap\de\phi=g^{ij}\partial_i\dd_j(\de\phi)~.\nonumber
\eea
Here an over-dot is a derivative w.r.t.~physical time $t$ and
$\al$, $\beta$ and $c_s^2$ are functions of $t$. In the case of the k-essence
field, from the action (\ref{action}) one finds~\cite{ke2}   
\be\label{cs}
c_s^2 = \frac{p'}{2Xp''+p'}~ \quad \quad '=\frac{d}{dX}~.
\ee
If for some time $c_s^2>1$, the fluctuations $\de\phi$ (or equivalently 
the Bardeen potential) propagate faster than the speed of light and
are therefore acausal. Indeed, equation (\ref{fluc}) can be rewritten as
\be\label{cone}
(G^{-1})^{\mu\nu}\partial_\mu(\de\phi)\partial_\nu(\de\phi)+\al\dot{\de\phi} + \beta\de\phi=0
\ee
where we have defined 
\be
(G^{-1})^{\mu\nu}=g^{\mu\nu}-(1-c_s^2)g^{ij}\de^{\mu}_i\de^{\nu}_j
\ee
as the inverse of the metric which governs the propagation of the
k-essence field. The characteristic cones of (\ref{cone}) are given 
by $(G^{-1})^{\mu\nu}$ ~\cite{CH}, and the rays by the metric 
\be
G_{\mu\nu}=g_{\mu\nu}+ \frac{1-c_s^2}{c_s^2}g_{ij}\de_{\mu}^i\de_{\nu}^j~.
\ee
If $c_s^2>0$, $G_{\mu\nu}$ is Lorentzian \cite{rendall}. However, 
if we consider a vector $n^\mu$ lying on the light cone defined by the
Einstein metric $g_{\mu\nu}$, we have
\be
G_{\mu\nu}n^\mu n^\nu=\frac{1-c_s^2}{c_s^2} g_{ij}n^in^j~.
\ee
Since $g_{ij}n^in^j <0$, $c_s^2>1$ implies $G_{\mu\nu}n^\mu n^\nu>0$, i.e. 
$n^{\mu}$ is time-like with respect to $G_{\mu\nu}$.
Therefore, the characteristic cone given by $G_{\mu\nu}$ is wider than
the light cone of causality defined by $g_{\mu\nu}$. 

But a k-essence field value at some event $q_0= (\eta_0,\bx_0)$ can be affected
by the values of all points inside the past characteristic
cone defined by $G_{\mu\nu}$ \cite{3femmes}. Let us now consider a point $q_1= (\eta_1,\bx_1)$
which is inside the past characteristic cone, but outside the past
light cone of $q_0$. Since $c_s^2>1$ such points exist and, in general,
the field value at $q_1$ influences the value at
$q_0$. However, since $q_1$ is outside the past light cone of
$q_0$, the distance $q_0-q_1$ is space-like and there exists a boost such
that the boosted event $q_1'$ is in the future of $q_0'$. 
In other words, the value of the field at $q_0'$ can be affected by its values in
the future; an evident a-causality. This is the well known consequence
of relativity. Whenever an event in $q$ is affected by something
outside its past light cone (defined by the metric which determines
causality, i.e.~the propagation of light and other standard model
particles), the present is affected by the future in some boosted 
reference frame. In order for causality to be
respected, it is therefore not enough that the k-essence
field propagates inside the light cone of $G_{\mu\nu}$ (as suggested
in \cite{picon}). If this cone is wider than the one of the Einstein
metric, this leads to superluminal propagation of the k-essence field
perturbations which are not acceptable.


Similar arguments in the more complicated case of multi-component
fields have led Velo and Zwanziger to the exclusion of generic higher
spin theories~\cite{VZ}. On the other hand, Gibbons \cite{Gib} has
analyzed the tachyon in the effective field model proposed by Sen
\cite{Sen}, and by the same argument has concluded that ``the tachyon
is not a tachyon'', because the characteristic cone of the tachyon lies inside
or on the light cone of the Einstein metric. The tachyon is unstable
but it does not violate causality.
The causality argument is also at the basis
of Ref.~\cite{AA} where it is used to exclude certain Lagrangians
as possible low energy approximations of a sound high energy field
theory. There it is also shown that this argument is not alleviated if
we allow the high energy theory to be a string theory. 
Therefore, the fluctuation equation (\ref{fluc}) leads to
acausalities if $c_s^2$ becomes larger than one (for an alternative view see \cite{bruneton}).

In the rest of the letter, we show that this is exactly what happens in
the case of successful k-essence models. We first present two examples from the
literature and then formulate a general proof showing that for a
successful k-essence model $c_s^2$ must become larger than one for some time. 

The equation of motion of the k-essence field is derived from the action
(\ref{action}). In order to have tracking solutions one must require
$K(\phi)=1/\phi^2$ \cite{ke+}. Moreover, 
to describe the dynamics of the k-essence field it is useful to consider 
the new variable $y\equiv 1/\sqrt{X}$ and introduce the function $g(y)
= p(X)/\sqrt{X}$.  In the new variables the energy density becomes 
$\rho_k=2X\frac{\partial P}{\partial X}-P = -g'/\phi^2$. Here a prime
denotes differentiation with respect to $y$.  
Since $\rho_k$ has to be positive, $g$ is monotonically decreasing,
$dg/dy=g'<0$. Other useful relations are~\cite{ke1} 
\bea 
w_k &=&\frac{P_k}{\rho_k} =\frac{p}{2Xp'-p}= \frac{-g}{yg'} \quad \mbox{ and } \quad \\
c_s^2 &=&\frac{p'}{2Xp''+p'} = \frac{g-g'y}{g''y^2}\label{cs_y}
\eea
(note that a prime on $p$ stands
for derivatives w.r.t.~$X$ while a prime on $g$ indicates
derivatives w.r.t.~$y$). The stability condition $c_s^2>0$ requires
$g''>0$ so that $g$ is convex~\cite{ke+}. 

In a Friedman universe with $H^2 = \rho_{\rm tot} = \rho_r +\rho_d
+\rho_k$  and $\Om_k=\rho_k/\rho_{\rm tot}$ ($\rho_r$ being the energy
density of  radiation and $\rho_d$ that of pressure-less matter, dust),
the k-essence equation of motion can be written in the form~\cite{ke1} 
\bea\label{eqmot}
\dot y &=&\frac{3(w_k(y)-1)}{2r'(y)}\left[ r(y)-\sqrt{\Om_k}\right] \\
\dot\Om_k &=& 3\Om_k(1-\Om_k)(w_m - w_k(y))~.\label{eqmot2}
\eea    
Here a dot indicates the derivative w.r.t $N=\ln(a)$, where $a$ is the
scale factor, $w_m$ is the ratio 
\bea
w_m &=&\frac{1}{3}\frac{\rho_r}{\rho_r+\rho_d}~, \quad \mbox{ and}
\\
\label{r(j)}
r(y) &\equiv& \frac{3}{2\sqrt{2}}\sqrt{-g'}(1+w_k)y =
\frac{3}{2\sqrt{2}}\frac{(g-g'y)}{\sqrt{-g'}}~. 
\eea
Fixed points, $y_f$ of the above system of equation are given by
$r^2(y_f) =\Om_k =$ constant, and
either $w_k(y_f)=w_m$, or $\Om_k=0$ or
$\Om_k=1$. These either are stable or can be made stable with
small changes in the function $g$~\cite{ke+}. 

The evolution of the universe 
drives the k-essence field from one fixed point to another. 
At early times, within a suitable range of initial conditions,
k-essence quickly approaches the radiation fixed point 
$y_r$. In order not to violate the nucleosynthesis bound one requires
$r(y_r)=\sqrt{\Om_k} \lesssim 0.1$. When the universe becomes matter
dominated, the radiation fixed point is lost and the k-essence energy
density decreases rapidly until the de Sitter attractor, $y_s$ with
$  0 \simeq \Om_k =r(y_s)\ll r(y_r)$ is reached.
From there, the field evolves to the k-essence attractor $y_k$ with
$r(y_k)=\sqrt{\Om_k}\simeq 1$ and $-1<w_k(y_k)<0$ or, if this
attractor does not exist, it evolves on to the dust attractor $y_d$
with $w_k(y_d)=w_m=0$. 
At present, the field
is on its way from the de Sitter fixed point up to either the k-essence
or dust attractor.

\begin{figure}[ht]
\centerline{\epsfig{figure=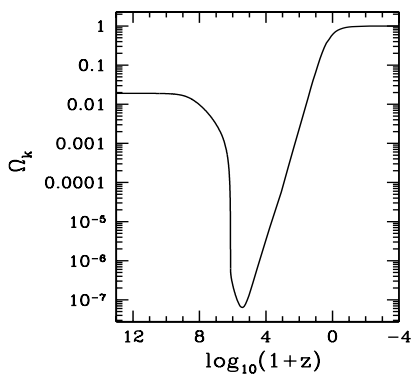,width=5.6cm,height=5.1cm}}
\caption{ \label{fig:Om} The ratio of k-essence to the total energy
  density $\Om_k$ as function of   
$1+z$ for the example (\ref{example_1}).
}
\end{figure}

\begin{figure}[ht]
\centerline{\epsfig{figure=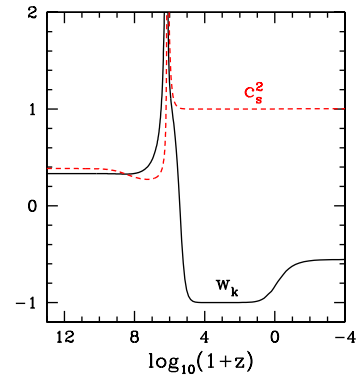,height=5.1cm}}
\caption{ \label{fig:cs} The equation of state parameter $w_k$  and
  sound velocity $c_s^2$ as functions of $1+z$ for the example
  (\ref{example_1}). 
}
\end{figure}

Examples of Lagrangians $P(\phi,X)$ that can be found in the
literature are~\cite{ke1,ke+} 
\be
P(\phi,X) = \frac{1}{\phi^2}\Big(-2.01+2\sqrt{1+X}
+3\cdot10^{-17}X^3-10^{-24}X^4 \Big) \label{example_1}  
\ee
and 
\bea
P(\phi,X) &=& \frac{1}{\phi^2}\Big(-2.05+2\sqrt{1+f(X)}\Big) \quad
\mbox{ where } \quad \nonumber\\ 
f(X)&=&X-10^{-8}X^2+10^{-12}X^3-10^{-16}X^4 \nonumber\\
&&+10^{-20}X^5-10^{-24}X^6/2^6~.
\label{example_2}
\eea
The evolution of interesting physical quantities for the Lagrangian
(\ref{example_1}) are shown  
in Figs.~\ref{fig:Om} and~\ref{fig:cs}. Example (\ref{example_2})
behaves similarly. In these examples, k-essence evolves to the final stage of
k-essence domination. 

Fig.~\ref{fig:cs} shows that $c_s^2$ becomes larger than $1$ during the
evolution from the radiation fixed point to the de Sitter fixed point and remains
slightly larger than $1$ in the future, when k-essence reaches the
k-attractor. 

In Ref.~\cite{AA}, it is shown that for small values of $X$ superluminal
propagation of perturbations around a non-trivial background is related to the
sign of the coefficient in front of the leading higher-dimension
operator in the 
Lagrangian: the absence of superluminal propagation requires a positive
coefficient. In both examples, expansion for small $X$ leads to 
$p(X)=a+bX+dX^2+\mathcal{O}(X^3)$, with $d<0$. For small $X$,
$p''(X)=2d$ is negative, and Eq.~(\ref{cs}) hence gives $c_s^2>1$. In
both examples, the theory is acausal. 
We note however that no problem arises in k-inflation, where the
coefficient in front of the term $X^2$ in the Lagrangian is
positive~\cite{ke2}.


We now demonstrate that $c_s^2>1$ is mandatory in every k-essence model
that aims to solve the coincidence problem and leads to accelerated
expansion of the universe today. From Eq.~(\ref{cs_y}),  $c_s^2>1$ is
equivalent to $g''y^2<g-g'y$. Using 
\be
 w_k = \frac{-g}{yg'}~,\label{wk} \quad
w'_k = \frac{gg'+gg''y-g'^2y}{(g'y)^2}~, 
\ee
and remembering that $g'<0$, 
we conclude that $w_k>1$ is equivalent to $g+yg' >0$ and $w_k'<0$ is
equivalent to $gg'+gg''y-g'^2y<0$. If both
these conditions are fulfilled,
we necessarily have $ 0> gg'+gg''y-g'^2y>-g'(-g+g''y^2+g'y)$. In the
last unequal sign we have used $g>-yg'$ and $g''> 0$. Since $g'<0$
this implies $g''y^2<g-g'y$ and therefore $c_s^2>1$.

We now show that such a situation always arises in k-essence
models which solve the coincidence problem. We first consider the
evolution of k-essence  from the 
radiation fixed point $y_r$ to the
k-essence fixed point $y_k$. We then show that the same arguments hold if 
the k-attractor is replaced by a late dust-attractor. 

We remind that, at a fixed point, $r(y_f)=\sqrt{\Om_k}$. 
Moreover one always has $y_k>y_r$ since $g$
is monotonically decreasing, and $w_k(y_k)<0$, hence $g(y_k)<0$ while
 $g(y_r)>0$ (see Eq.~(\ref{wk})). 
Eq.~(\ref{r(j)}) gives
\be \label{r'}
\frac{dr}{dy} = \frac{3}{2\sqrt{8}}\frac{g''y}{\sqrt{-g'}}(w_k(y)-1)~,
\ee
and since $g''>0$, $r(y)$ can increase only if
$w_k>1$. Since $r(y_k) > r(y_r)$, $r$ has to increase 
from $y_r$ to $y_k$. 
This implies that there exists an interval $y_0<y<y_2$, with
$y_r<y_0<y_2<y_k$, in which $w_k(y)>1$. Since $w_k(y_k)<0$ we can choose
without loss of generality $w_k(y_2)=1$.  
For some part of the interval, say in $]y_1,y_2[$, $w_k$
has to decay: $w_k'<0$. Therefore, both
conditions necessary for $c_s^2>1$ are satisfied in $]y_1,y_2[$.
In other words, between two points $y_a<y_b$
with  $r(y_a)<r(y_b)$ and $w(y_a)<1,~ w(y_b)<1$ there exists necessarily an
interval with $c_s^2(y)>1$.

If the k-attractor is replaced by a late dust-attractor, the situation is 
alike. Indeed, since $w_k(y_d)=0$ hence $g(y_d)=0<g(y_r)$ we must have
$y_r<y_d$. Furthermore, in order to have a period of accelerated expansion with
$w_k<-1/3$,  we need $r(y_d)>r(y_r)$. If $r(y_d)<r(y_r)$, the accelerating
phase is avoided because the k-essence fluid is attracted immediately to the 
dust-attractor after matter-radiation equality~\cite{ke+}. 
Therefore $r'(y)$ must be positive in an interval between $y_r$
and $y_d$ and the demonstration above holds also in this case.

The only behavior relevant for this result is the existence of a
radiation tracker which goes over into an accelerating phase
with $w_k<-1/3$ and a
relatively large value of $\Om_k$, as we observe it today. During such
a phase $\Om_k$ must increase according to~(\ref{eqmot2}) and, if it is to
reach a fixed point $y_f$ with $\Om_k =r^2(y_f)$ and $y_f>y_r$, the
function $r(y)$ is bound to increase somewhere, which is sufficient for
$c_s^2>1$ as we have shown above.
On its way
from the radiation fixed point, $w_k(y_r)=1/3,~ \Om_k =r^2(y_r)\ll 1$ to
the k-essence  fixed point with $w_k(y_k) <0,~\Om_k =r^2(y_k) \simeq 1$ (or to
the dust fixed point with $w_k(y_d)=0,~\Om_k=r^2(y_d)>r^2(y_r)$) the
k-essence fluid has to pass through an interval where $c_s^2>1$.

The fact that $w_k$ has to be larger than $1$ in some interval for a
successful k-essence model was already pointed out in
Ref.~\cite{ke+}. This means that there exists observers which see an
energy flow which is faster than the speed of light. But this does not
necessarily pose a problem for causality since the energy flow does not carry
information. However, $c_s^2$
represents the propagation velocity of the perturbations, at least in
the WKB limit which is always justified for large enough wave
numbers, and therefore it
really means that information can travel faster than light. 

We have shown that k-essence which has the capacity to play the role
of dark energy and, especially, to address the coincidence problem
cannot result as a low energy effective theory from some meaningful,
causal high energy field theory, because it necessarily undergoes phases
where $c_s^2>1$. In the examples presented here, 
$c^2_s>1$ also at late times. This means that
k-essence models which solve the coincidence problem are ruled out as
serious candidates for dark 
energy. However, the k-essence model proposed in Ref.~\cite{scherer},
which does not solve the coincidence problem, does not have
$c_s^2>1$. Also the form of the Lagrangian needed for successful
k-inflation usually does not suffer from this problem and has causally
propagating fluctuations.

 This work is supported by the Swiss NSF.

\end{document}